\documentclass[a4paper,11pt]{article}
\usepackage{pos}

\usepackage{epstopdf}
\def\bea{\begin{eqnarray}}
	\def\eea{\end{eqnarray}}

\def\pp{\mbox{$p$-$p$}}

\def\pa{\mbox{$p$-A}}

\def\pbpb{\mbox{Pb-Pb}}
\def\ppb{\mbox{$p$-Pb}}

\def\aa{\mbox{A-A}}
\def\nn{\mbox{N-N}}

\def\pt{$p_t$}

\def\yt{$y_t$}

\def\nch{$n_{ch}$}
\def\mmpt{$\bar p_t$}

\title{Space-time Geometry of Small and Large Collision Systems}
 \ShortTitle{Space-time Geometry of Small and Large Collision Systems}

\author[a]{Thomas A. Trainor}
\affiliation[a]{University of Washington, Seattle, USA}
\emailAdd{ttrainor99@gmail.com}
\abstract{
Identified-hadron spectra from 2.76 TeV Pb-Pb and $p$-$p$ collisions are analyzed via a two-component (soft + hard) model (TCM) of hadron production in high-energy nuclear collisions. The object of study is evidence for jet suppression in small and large collision systems. Conventional methods include Pb-Pb centrality determination via classical Glauber model and evidence for high-$ p_t$ suppression sought via spectrum ratio $R_\text{AA}$. Previous $p$-Pb studies questioned the validity of the classical Glauber model. In the present study A-A geometry is determined instead via ensemble-mean $\bar p_t$ data. Based on certain features of Pb-Pb spectra the validity of the factorization assumption is also questioned. The entire jet contribution is therefore treated without factorization in ratio to a \mbox{$p$-$p$} spectrum model as reference. These new results indicate that {\em exclusivity} (a nucleon may only interact with one nucleon ``at a time'') and time dilation (experienced by participant partons) play an essential role in jet production not incorporated in Glauber model or hard-component factorization. The combination determines an effective number of N-N collisions per participant nucleon given specific Pb-Pb centrality: multiple collisions if associated with low-$x$ (slow) partons, a single collision if associated with high-$ x$ (fast) partons experiencing strong time dilation. The effect on parton fragment (jet) distributions on $p_t$ may be misinterpreted as jet suppression, but is similar to projectile-proton fragment distributions on pseudorapidity from fixed-target $p$-A experiments where low-$\eta$ densities scale with A while high-$\eta$ densities are consistent with $p$-$p$ collisions. $p$-Pb and Pb-Pb spectra similarly analyzed reflect the same physics given different geometries. Actual jet suppression related to QGP formation is not evident.
}

\FullConference{Proceedings of the Corfu Summer Institute 2025 "School and Workshops on Elementary Particle Physics and Gravity" (CORFU2025)\\
27 April - 28 September, 2025\\
Corfu, Greece\\}

\begin{document}
\maketitle

%\vspace{10pt}
%\noindent\rule{\textwidth}{1pt}
%\tableofcontents\thispagestyle{fancy}
%\noindent\rule{\textwidth}{1pt}
%\vspace{10pt}

%%%%%%%%%%%%%%
\section{Introduction} \label{intro}
%\label{sec:intro}

The relativistic heavy ion program was initiated to attempt formation of a quark-gluon plasma or QGP. Expected indicators for such formation included observation of a flowing particle source (radial flow), azimuth modulation of radial flow as elliptic flow and modification of jet formation due to a dense QCD medium (jet quenching). This talk considers evidence for the last, especially as it relates to the space-time geometry of \aa\ collision systems.

Collision geometry (impact parameter, participant nucleon number) is conventionally determined by classical Glauber Monte Carlo. However, recent studies of small asymmetric collision systems (e.g.\ \ppb\ collisions)~\cite{ppbpid} indicate that Glauber models overestimate the number of participant nucleons in \aa\ collisions by up to a factor 3 (Sec.~\ref{glauber})~\cite{tomglauber}. An alternative approach employs ensemble-mean \mmpt\ data to infer hard/soft ratio $x\nu$ (Sec.~\ref{tcmgeom}).

Spectrum structure is analyzed with a two-component (soft+hard) model (TCM) that accurately separates soft (projectile-nucleon) fragments from hard (scattered-parton) fragments (Secs.~\ref{tcm}, \ref{standard}). Interpretable study of possible jet modification relies on complete isolation of the spectrum hard component. Certain features of spectrum hard components so isolated call into question some basic assumptions, especially factorizability of the hard component based on estimates of the number of \nn\ collisions $N_{coll}$ or {\em binary} collisions $N_{bin}$ (Sec.~\ref{factor}).

Resolution of apparent contradictions emerges based on {\em exclusivity}~\cite{tomexclude} and time dilation as relates to \nn\ collisions and parton momentum fraction $x$ (Sec.~\ref{dilation}). Reformulation of TCM hard-component ratio measure $r_{AA}$ leads to a new presentation format in which jet fragments are apparently not suppressed. Instead, jet fragment production is limited based on quantum mechanics and relativistic time dilation. This talk summarizes results recently reported in Ref.~\cite{tompbpb}.

%%%%%%%%%%%%%%
\section{Conventional spectrum presentation format} \label{convention}

\begin{figure}[h]
	\includegraphics[width=1.46in,height=1.41in]{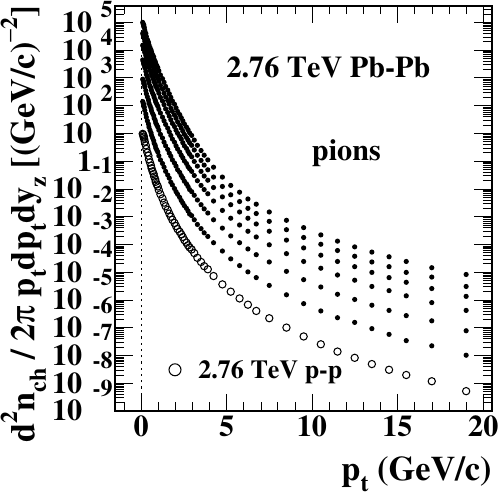}
	\includegraphics[width=1.46in,height=1.4in]{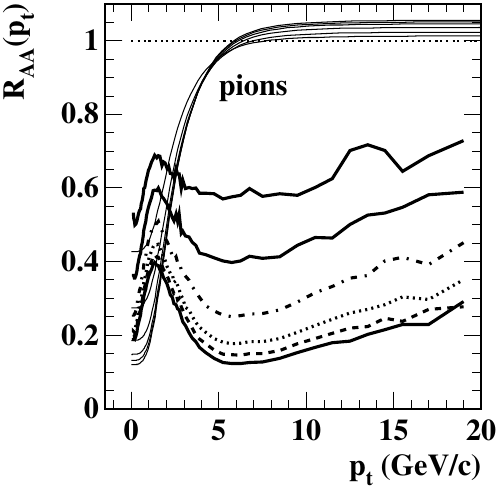}
	\includegraphics[width=1.46in,height=1.46in]{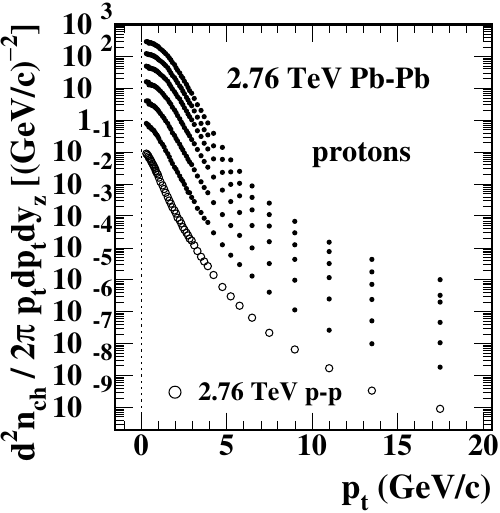}
	\includegraphics[width=1.46in,height=1.4in]{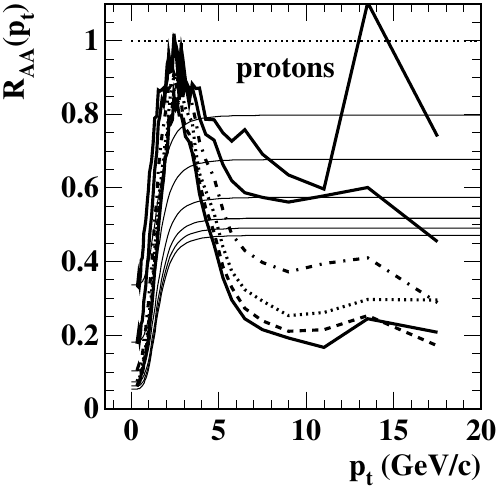}
	\caption{\pt\ spectra and $R_{AA}$ for
		(a,b) pions and
		(c,d) protons in conventional format.
	}
	\label{fig1}     
\end{figure}

Figure~\ref{fig1} shows pion and proton \pt\ spectra and ratios $R_\text{AA}$ for 2.76 TeV \pbpb\ collisions in a conventional plotting format~\cite{alicepbpbpidspec}. A rescaled spectrum ratio, defined as 
\bea \label{raa}
R_\text{AA}(p_t) \equiv (1/N_{bin}) \bar \rho_\text{0AA}(p_t) / \bar \rho_\text{0pp}(p_t),
\eea
is expected to indicate jet modification by reduction of $R_\text{AA}$ below unity at higher \pt.  That format and measure are based in part on certain assumptions: Most produced particles come from a single source identified as a QGP (flowing bulk medium). Jets are a ``high-\pt'' phenomenon. Projectile nucleons may interact simultaneously with multiple target nucleons (Glauber).

%%%%%%%%%%
%%%%%%%%%

The conventional format is based on questionable assumptions and results in suppression of critical information carried by spectrum data. Spectra for successive event classes are rescaled by powers of 2 (thus increasing vertical separation) and plotted  as individual points rather than curves. The combination limits precise differential comparison that might reveal details of spectrum evolution with \nch. Data are plotted vs linear \pt, thereby favoring a broad higher-\pt\ interval that exhibits little structure. The interval above 4 GeV/c that is favored includes about 5\% of all jet fragments. The great majority appear near 1 GeV/c (note the peak structures in panels b,d). The thin curves are linear-superposition references (what to expect if  there is no jet suppression) that depend strongly on hadron species. For instance, any proton suppression for most-central collisions in panel (d) should be measured relative to the lowest thin curve, not 1. The basis for generating rescale factor $1/N_{bin}$ in Eq.~(\ref{raa}) is questionable~\cite{tomglauber}.

%%%%%%%%%%%%%%
\section{Two-component $\bf p_t$ spectrum model} \label{tcm}

A two-component \pt\ spectrum model (TCM) was first inferred from 200 GeV \pp\ spectrum evolution with event \nch~\cite{ppprd}. No assumptions preceded the analysis. Inferred model elements were found to be required by data, their physical interpretations derived later by comparisons with QCD theory.
The basic \pp\ spectrum model may be expressed as
\bea
\bar \rho_0(p_t) \equiv d^2n_{ch}/p_t dp_t d\eta = \bar \rho_s \hat S_{0}(p_t) + \bar \rho_h \hat H_{0}(p_t,n_{ch}) ~~~~ \text{soft + hard}
\eea
representing participant nucleon dissociation (soft) plus jet fragments (hard).
For individual \nn\ collisions within \aa\ collisions $\bar \rho_x \rightarrow \bar \rho_{xNN}$. {\em Empirically-inferred} relation $\bar \rho_{hNN} \approx \alpha \bar \rho_{sNN}^2$, with hard/soft density ratio $x = \bar \rho_{hNN} / \bar \rho_{sNN} \approx \alpha \bar \rho_{sNN}$ and $\alpha \approx 0.01$, is an all-important element of the model.
For identified-hadron (PID) spectra
\bea \label{pidtcm}
\bar \rho_{0i(p_t,n_{ch})} = \bar \rho_{si} \hat S_{0i}(p_t) + \bar \rho_{hi} \hat H_{0i}(p_t,n_{ch}),
\eea
where for hadron species $i$ $\bar \rho_{si} = z_{si}(n_s)(N_{part}/2) \bar \rho_{sNN}$ and
$\bar \rho_{hi} = z_{hi}(n_s)N_{bin} \bar \rho_{hNN}$. The $z_{xi} \leq 1$ are fractional abundances for hadron species $i$ with $\bar \rho_{hi} / \bar \rho_{si} = \tilde z_i x \nu$ and $\tilde z_i  = z_{hi} / z_{si} \propto$ hadron mass. Those relations and parameter values are inferred from 5 TeV \ppb\ spectrum data in Refs.~\cite{ppbpid}.
Spectrum hard component $\bar \rho_{h} \hat H_{0}(p_t)$ for NSD \pp\ collisions is consistent with QCD collinear factorization and measured properties of minimum-bias jets~\cite{fragevo,jetspec2}.

Geometry parameters $N_{part}$ and $N_{bin}$ plus densities $\bar \rho_{xNN}$ must be obtained from a collision model or other source of information. A conventional approach employs a classical Glauber Monte Carlo model to obtain AB geometry information as discussed in the next section.

%%%%%%%%%%%%%%
\section{Classical Glauber Monte Carlo geometry and alternative} \label{glauber}

A Glauber Monte Carlo is based on a model of colliding projectiles in terms of interpenetrating spheres. Two nucleons are said to have interacted or collided based on an inelastic cross section measured in an {\em isolated} \nn\ system. If an \nn\ impact parameter is less than a corresponding limit an interaction has taken place defining {\em participant nucleons} and \nn\ {\em binary collisions}. 

\begin{figure}[h]
	\includegraphics[width=1.46in,height=1.4in]{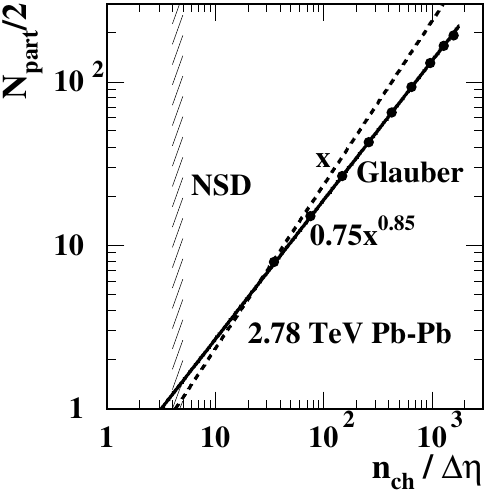}
	\includegraphics[width=1.46in,height=1.42in]{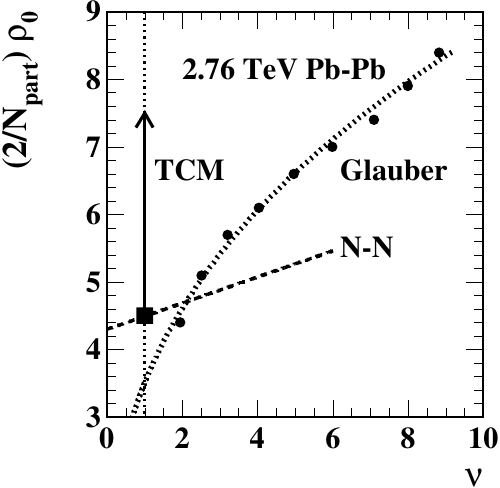}
	\includegraphics[width=1.46in,height=1.42in]{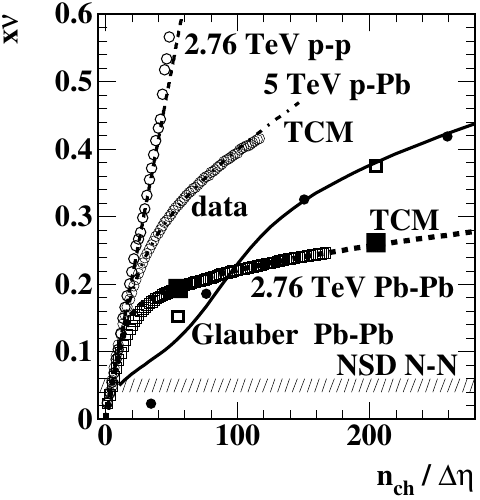}
	\includegraphics[width=1.46in,height=1.42in]{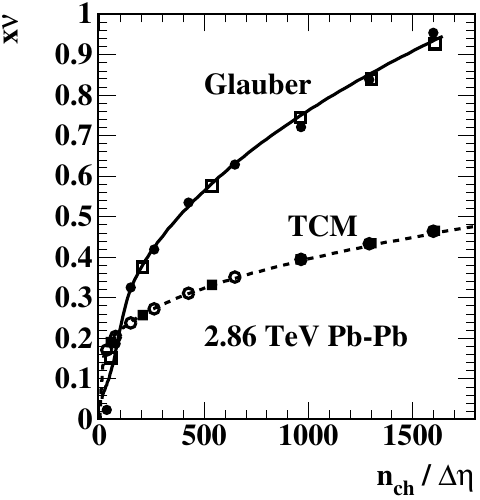}
	\caption{
		(a) Glauber $N_{part}/2$ vs $\bar \rho_0$,
		(b) Glauber $(2/N_{part}) \bar \rho_0$ vs $\nu$,
		(c) TCM product $x\nu$ inferred from \mmpt\ data,
		(d) Extrapolation of TCM $x\nu$ trend to large \pbpb\ \nch.
	}
	\label{fig2}     
\end{figure}

Figure~\ref{fig2} (a) shows $N_{part}$ vs $\bar \rho_0$ (points) from Refs.~\cite{pbpbcent,alicepbpbyields}. The trend is a power law $N_{part}/2 \propto \bar \rho_0^{0.85}$ (solid line). Panel (b) shows $(2/N_{part}) \bar \rho_0$ (results from Ref.~\cite{alicepbpbyields}) vs $\nu$ where $\nu \equiv 2N_{bin} / N_{part}$. A second Glauber power-law relation is $N_{bin} \propto N_{part}^{1.45}$~\cite{tompbpb}. Combining those results gives $(2/N_{part}) \bar \rho_0 \propto \nu^{0.40}$ (bold dotted curve). The solid square corresponds to NSD \pp\ collisions. It is notable that the Glauber trend substantially misses the \pp\ limiting case. Detailed study of centrality issues for 5 TeV \ppb\ collisions indicates that assumptions for \nn\ collisions within A-B collisions are inconsistent with data, in part due to exclusivity~\cite{ppbpid,tomexclude}.

%%%%%%%%%%
%%%%%%%%%

In Ref.~\cite{ppbpid} centrality for 5 TeV \ppb\ collisions is addressed. The \ppb\ system is amenable to exact determination of collision geometry based on ensemble-mean \mmpt\ data. Panel (c) shows hard/soft ratio $x\nu$ obtained by inversion of the relation $\bar p_t = (\bar p_{ts} + x\nu \bar p_{th})/(1 + x\nu)$ where fixed mean values $\bar p_{ts}$ and $\bar p_{th}$ are obtained from TCM model functions $\hat S_0(p_t)$ and $\hat H_0(p_t)$ required to describe spectra. It is notable that three collision systems exhibit identical $x\nu$ trends below a transition point at $\bar \rho_0 \approx 15$ which is one manifestation of exclusivity~\cite{tomexclude}. Panel (d) shows extrapolation of the TCM $x\nu$ trend to the full centrality range of \pbpb\ collisions for the present study. It is especially notable that the Glauber results dramatically disagree with what is inferred from \mmpt\ data, for example a factor two difference at right in panel (d).

%%%%%%%%%%%%%%
\section{TCM-derived Pb-Pb  geometry} \label{tcmgeom}

\begin{figure}[h]
	\includegraphics[width=2.92in,height=1.4in]{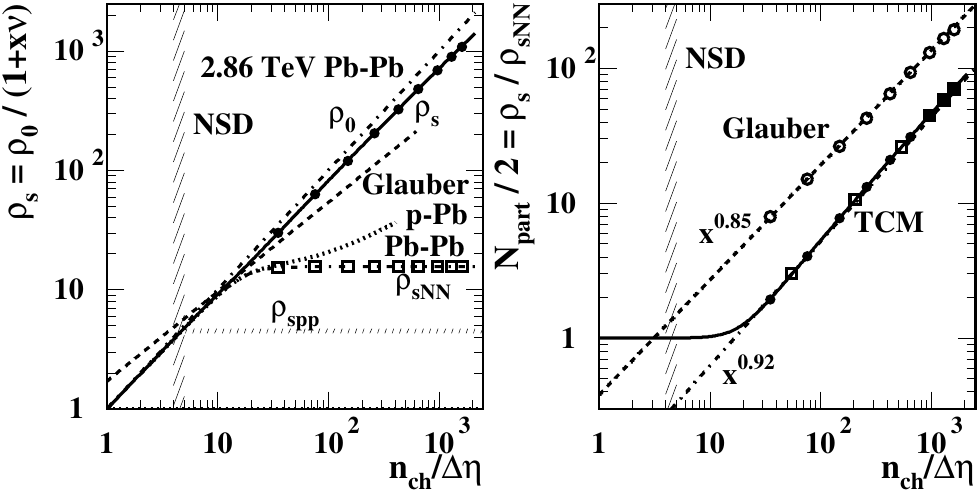}
	\includegraphics[width=2.92in,height=1.42in]{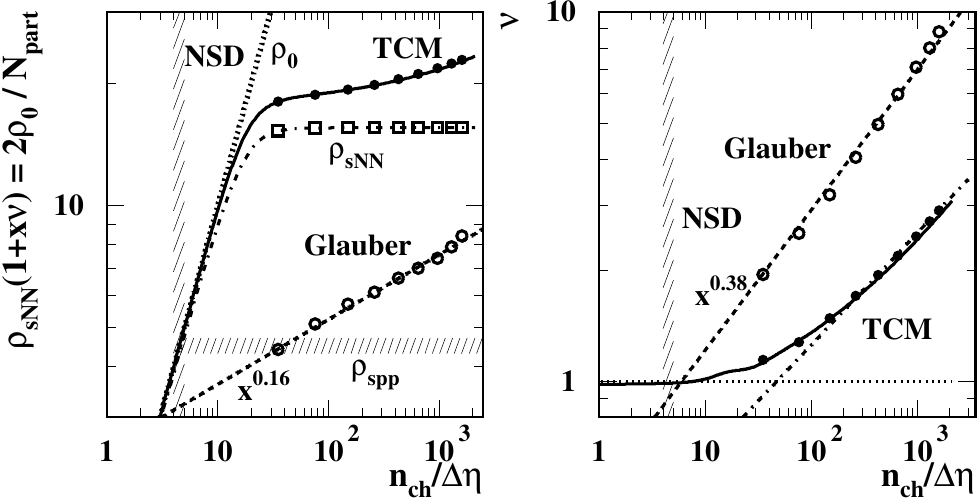}
	\caption{TCM 
		(a) $\bar \rho_s$ vs $\bar \rho_0$
		(b)  $N_{part} / 2$ vs $\bar \rho_0$
		(c)  $(2/N_{part}) \bar \rho_0$ vs $\bar \rho_0$
		(d)  $\nu$ vs $\bar \rho_0$
	}
	\label{fig3}     
\end{figure}

Using the $x\nu$ trend inferred from \mmpt\ data as described above, other \pbpb\ geometry parameters are derived following the procedure in Ref.~\cite{ppbpid}. Figure~\ref{fig3} (a) shows soft-component charge density $\bar \rho_s$ (solid points and curve) derived from $\bar \rho_s = \bar \rho_0/(1 + x\nu)$. As for \ppb\ \nn\ soft density $\bar \rho_{sNN}$ (dotted curve) \pbpb\ $\bar \rho_{sNN}$ (lower dash-dotted curve and open squares) follows $\bar \rho_s $ up to a transition point near $\bar \rho_0 \approx 15$ and then proceeds with a much-reduced slope, here assumed zero for \pbpb. In panel (b) the participant-pair number $N_{part}/2 = \bar \rho_s  / \bar \rho_{sNN}$ (solid curve and points) follows from Sec.~\ref{tcm}. Panel (c) shows $2 \bar \rho_0 / N_{part} = \bar \rho_{sNN}(1+x\nu)$ (solid curve and points) that may be contrasted with the Glauber version in Fig.~\ref{fig2} (b) (and open circles in this panel) from Ref.~\cite{alicepbpbyields}. Panel (d) shows corresponding parameter $\nu$.

%%%%%%%%%%
%%%%%%%%%

There are two fundamental aspects to the large differences between Glauber and TCM geometry: (i) Peripheral A-B collisions are equivalent to single \nn\ collisions over a substantial \nch\ interval due to exclusivity (see equivalence of three collision systems over that interval in Fig.~\ref{fig2}, c). A second interaction is not possible until the \aa\ overlap space-time volume is sufficiently large, corresponding to the transition point at $\bar \rho_0 \approx 15$ (three times the NSD value 4.55)~\cite{tompbpb}. As a result, $N_{part}/2 \approx N_{bin} \approx 1$ within that interval. (ii) Also as a result, $\bar \rho_{sNN}$ in panel (c) increases to three times the NSD value (hatched) and then becomes slowly varying (dash-dotted curve) whereas the Glauber trend is dominated by $N_{part} \propto \bar \rho_0^{0.85}$ as in Fig.~\ref{fig2} (a).

%%%%%%%%%%%%%%
\section{Standard TCM spectrum analysis} \label{standard}

\begin{figure}[h]
	\includegraphics[width=2.92in,height=1.4in]{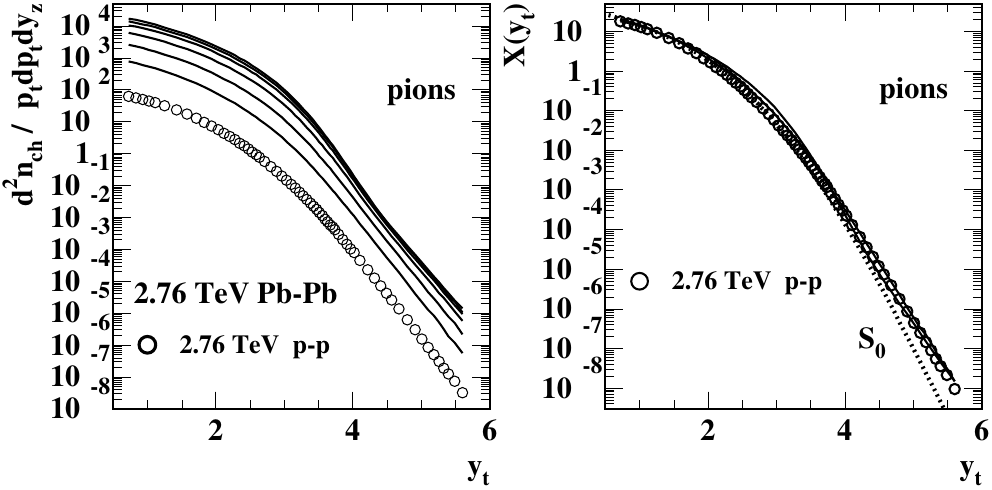}
	\includegraphics[width=2.92in,height=1.42in]{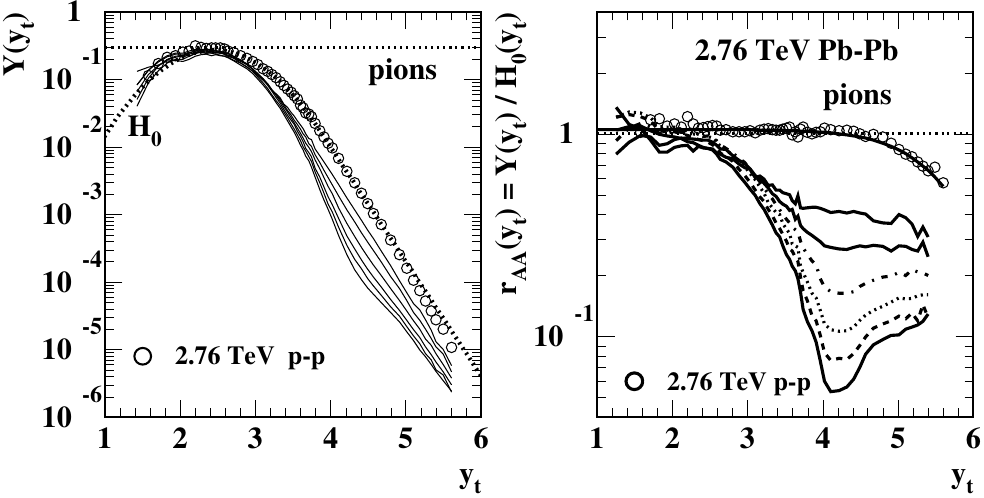}
	\caption{TCM analysis for pions with
		(a) \pt\ spectrum $\bar \rho_{0i}(y_t)$,
		(b) soft-rescaled spectrum $X_i(y_t)$,
		(c) hard-rescaled hard component $Y_i(y_t)$,
		(d) ratio $r_{AAi} = Y_i(y_t) / \hat H_{0i}(y_t)$
	}
	\label{fig4}     
\end{figure}

Figure~\ref{fig4} (a) shows published \pt\ spectra (solid curves) for pions from 2.76 TeV \pbpb\ collisions, \pt\ densities as defined on the $y$ axis {\em without rescaling}. Also shown are corresponding \pp\ spectra (open circles). It is notable in this {\em un}rescaled format that the separations between different event classes are quite irregular, especially the large gap between most-peripheral \pbpb\ and \pp. As shown below, that spacing of event classes negatively impacts physical interpretation. The independent variable {\em transverse rapidity} $y_t = \ln[(m_{ti}+ p_t)/m_i]$ helps to resolve spectrum structure especially at lower \pt\ {\em where most jet fragments appear}~\cite{ppprd,fragevo,jetspec2}.

%%%%%%%%%%
%%%%%%%%%

Figure~\ref{fig4} (b) shows soft-rescaled spectra as $X_i(y_t) = \bar \rho_{0i}(y_t) / \bar \rho_{si}$ which ensures that all spectra coincide at low \pt, providing precise differential comparison of spectra. Soft-component model $\hat S_{0i}(y_t)$ is the dotted curve defined as the asymptotic limit of data spectra as $n_{ch} \rightarrow 0$.

Figure~\ref{fig4} (c) shows  $Y_i(y_t) = [X_i(y_t) - \hat S_{0i}(y_t)] / \tilde z_i x\nu$, the spectrum hard component (jet fragment distribution) with hard rescale by factor $\tilde z_i x\nu$ assuming nucleons {\em interact as monoliths}. The data may be compared with the hard-component model function $\hat H_{0i}(y_t)$ (dotted).

Figure~\ref{fig4} (d) shows  $Y_i(y_t) / \hat H_{0i}(y_t)$, a ratio similar to conventional $R_{AA}$ but including only jet-related particles over the entire \pt\ acceptance. Whereas $R_{AA}$ in Fig.~\ref{fig1} (b,d) emphasizes $p_t > 4$ GeV/c, panel (d) indicates that jet production {\em below} 4 GeV/c ($y_t \approx$ 4) is most interesting. Note that \pp\ data in (d) drop off at higher \yt\ for these data (note solid model curve, the same for each hadron species) which has not been observed for other \pp\ data~\cite{alicetomspec}.

%%%%%%%%%%%%%%
\section{Questioning hard-component factorization} \label{factor}

The (up to now) standard TCM analysis summarized in the previous section implicitly assumes that nucleons interact as monoliths. The emphasis has been on correct modeling of A-B collision geometry via \mmpt\ data (facilitated by \ppb\ studies) and precise separation of soft and hard components so as to obtain an accurate description of jet production and resulting fragments.
With geometry issues resolve via \ppb\ analysis and broad experience with a number of collision systems and energies, certain curiosities have persisted that were evident even at RHIC.

\begin{figure}[h]
	\includegraphics[width=1.46in,height=1.4in]{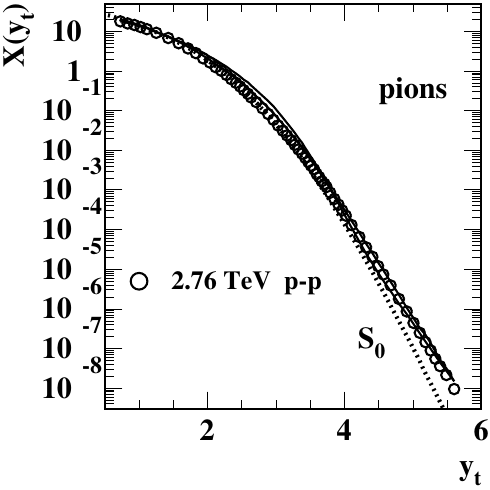}
	\includegraphics[width=1.46in,height=1.4in]{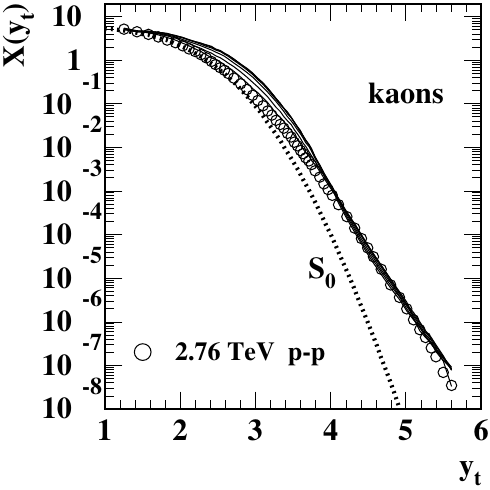}
	\includegraphics[width=2.92in,height=1.42in]{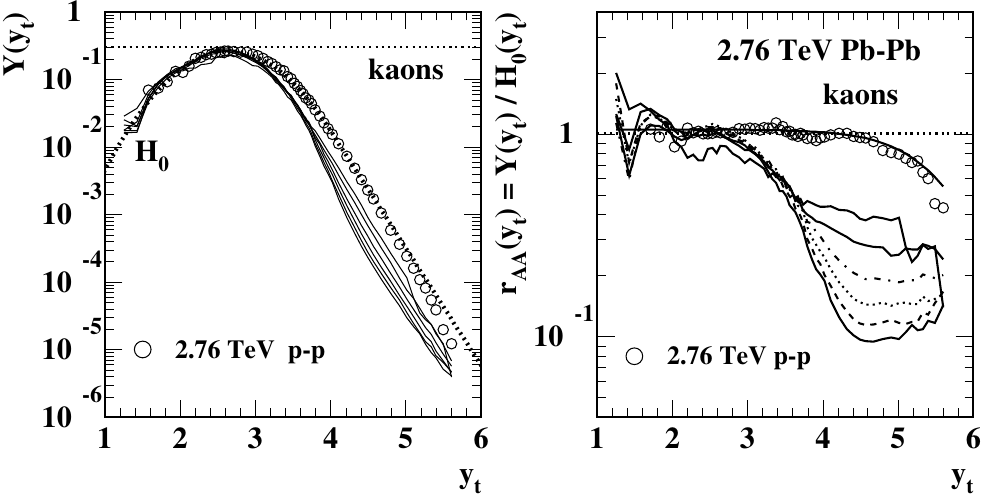}
	\caption{soft-rescaled spectrum $X_i(y_t)$ for
		(a) pions and
		(b) charged kaons,
		(c) hard-rescaled hard component $Y_i(y_t)$ and
		(d) ratio $r_{AA} = Y_i(y_t) / \hat H_0(y_t)$ for  kaons.
	}
	\label{fig5}     
\end{figure}

Figure~\ref{fig5} (a,b) shows spectra for pions and kaons in the soft-rescaled form $X_i(y_t)$ that by construction should coincide at lower \pt\ over all event classes but {\em also} similarly coincide with \pp\ spectra at higher \pt. If high-\pt\ spectrum evolution were due to jet suppression it seems a remarkable conspiracy that jet suppression should lead to such coincidence at high \pt. Note the much greater jet contribution for kaons in (b) than for pions in (a) since the {\em relative} jet contribution (hard-to-soft ratio measured by $\tilde z$) is proportional to hadron mass~\cite{pidpart1}.

%%%%%%%%%%
%%%%%%%%%

Figure~\ref{fig5} (c,d) show nominal evidence for jet suppression in a data/reference ratio format similar to $R_{AA}$ but here spanning the entire \pt\ acceptance down to zero \pt. The apparent suppression at higher \pt\ corresponds (at some level) to that indicted in Fig.~\ref{fig1} (b,d), but there is no corresponding suppression at lower \pt\ (which is effectively hidden by the $R_{AA}$ ratio). Note the quality of the  TCM {\em jet-related} data description even below$y_t \approx 2$  ($p_t \approx 0.5$ GeV/c).

Those results suggest that certain assumptions supporting the TCM (in common with other approaches) should be questioned. Hard-component factorization to infer an outcome {\em per \nn\ collision} (e.g.\ Fig.~\ref{fig4}, c) assumes that nucleons are in effect monolithic and behave similarly in any context. Thus, $N_{bin}$ should represent the number of nucleon-nucleon interactions as opposed to other processes. However, current assumptions, including those for the TCM, lead to paradoxical results as in Fig.~\ref{fig5} suggesting the need for an alternative narrative.

%%%%%%%%%%%%%%%%%%
\section{Exclusivity and Time Dilation} \label{dilation}

The large differences between classical Glauber model and A-B geometry inferred from \mmpt\ data may be traced to two elements -- exclusivity and relativistic time dilation -- affecting \nn\ collisions within A-B collisions. Exclusivity is manifested by the trends in Fig.~\ref{fig3} for which $N_{part}/2 \approx N_{bin} \approx 1$ over a significant \nch\ interval above NSD \pp\ collisions. The effect becomes apparent with \ppb\ collisions for which the geometry is exactly solvable based on \mmpt\ data~\cite{ppbpid} and is described as such in Ref.~\cite{tomexclude}. \ppb\ data imply that a projectile nucleon may only interact with one target nucleon ``at a time,'' where the dead time is at least one nucleon diameter. That principle leads to $N_{part}$ and $N_{bin}$ trends that come close to providing a self-consistent data description. However, ``at a time'' is not sufficiently well defined since the ``clock'' that measures time is not specified, and consequences for \pt\ dependence at higher \pt\ emerge as follows.

A second issue for the classical Glauber model is assumption of nucleons as monolithic structures that form the basis for counting. That assumption ignores the composite nature of nucleons as a system of partons with broad speed distribution and consequent {\em time dilation}. Low-$x$ (momentum fraction $x$) partons have fast clocks whereas high-$x$ partons have slow clocks. If an \nn\ interaction is carried by a low-$x$ parton (corresponding to low-\pt\ hadron production) the dead time may correspond to a nucleon diameter, whereas for high-$x$ partons (corresponding to high-\pt\ hadron production via jets) the dead time may exceed a Pb diameter. Thus, low-\pt\ hadrons may originate from multiple \nn\ interactions within an A-B collision whereas high-\pt\ hadrons may come from single \nn\ interactions. Glauber estimates $N_{coll}$ (number of \nn\ collisions) or $N_{bin}$ (number of {\em binary} collisions) may then be quite misleading.

%%%%%%%%%%%%%%%%%%
\section{Modified TCM -- Pb-Pb hard components} \label{pbpb}

By the above argument factorization as $Y_i(y_t) = [X_i(y_t) - \hat S_{0i}(y_t)] / \tilde z_i x\nu$ (spectrum hard component) is not valid since $x\nu$ is no longer well defined. In its place unfactorized hard component $Y'_i(y_t) \equiv X_i(y_t) - \hat S_{0i}(y_t) \leftrightarrow \tilde z_i x\nu \hat H_{0i}(y_t)$ is retained and compared to an established reference, being the hard-component model for corresponding \pp\ collisions. For species $i$
\bea
r_Y(y_t) = Y'_{AA}(y_t)/Y'_{NN}(y_t) = [\tilde  z x_{AA} \nu \hat H_{0AA}(y_t)] / [\tilde z x_{NN} \hat H_{0NN}(y_t)],
\eea
where the numerator at right is now the object of study. The numerator is the full spectrum hard component rescaled by the total soft density $\bar \rho_{si}$ as per quantity $X_i = \bar \rho_{0i} / \bar \rho_{si}$.

\begin{figure}[h]
	\includegraphics[width=1.46in,height=1.4in]{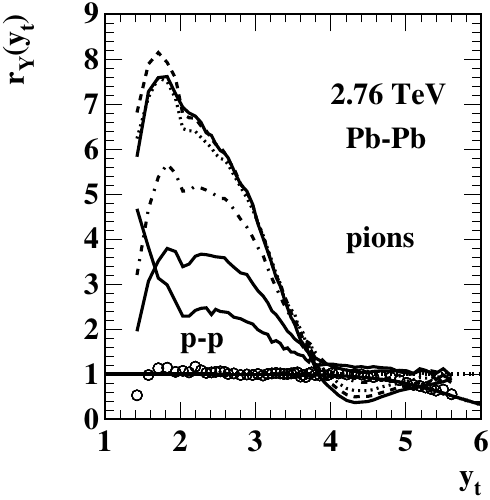}
	\includegraphics[width=1.46in,height=1.4in]{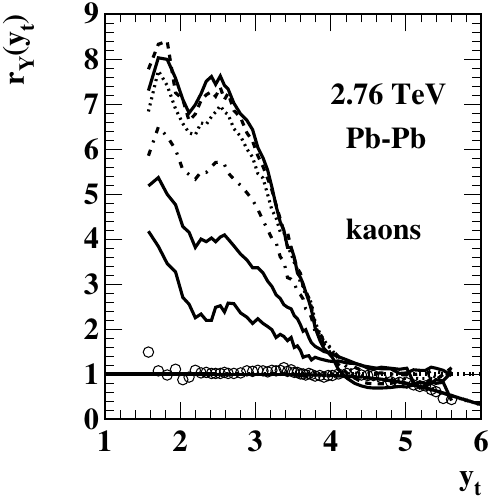}
	\includegraphics[width=2.92in,height=1.4in]{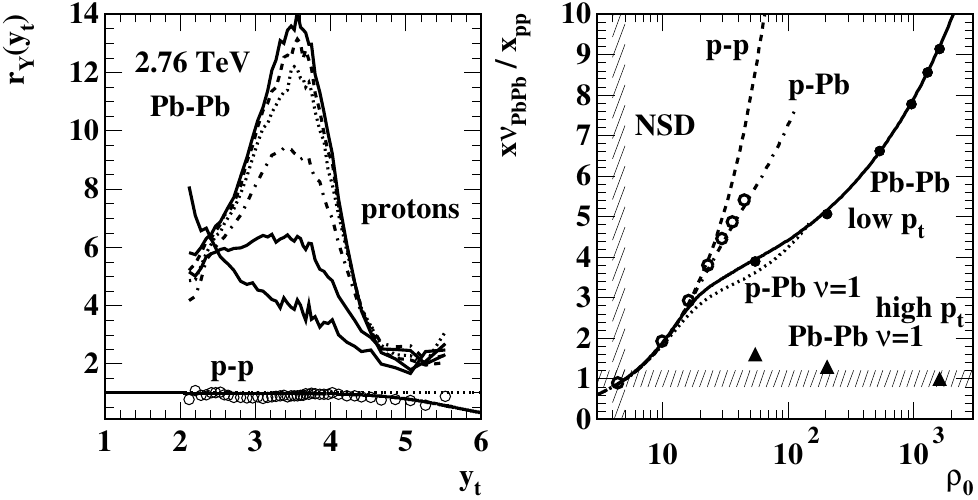}
	\caption{Revised spectrum ratio $r_Y(y_t)$ for \pbpb\ collisions and for
		(a) pions,
		(b)  kaons and
		(c) protons;
		(d) ratio $x\nu_{PbPb} / x_{pp}$ for \pp, \ppb\ and \pbpb\ collisions.
		\\  check ratio if xpbpb is not xNN -- why is transition at 3 and not 1?
	} 
	\label{fig6}     
\end{figure}

Figure~\ref{fig6} shows modified spectrum ratio $r_{Y}(y_t)$ for (a) pions, (b) charged kaons and (c) protons from 2.76 TeV \pbpb\ collisions. Given  the previous section one may interpret results as follows: Below \yt\ = 4 ($p_t < 4$ GeV/c) hadrons originate from lower-$x$ partons and correspond to projectile nucleons undergoing multiple \nn\ collisions within a \pbpb\ collision but constrained by exclusivity to a limited number of such  collisions. Above \yt\ = 4, hadrons from higher-$x$ partons correspond to participant nucleons constrained to single \nn\ collisions within a \pbpb\ collision because of a combination of exclusivity and parton time dilation.

%%%%%%%%%%
%%%%%%%%%%

Figure~\ref{fig6} (d) shows ratio $x_{PbPb} \nu / x_{pp}$ (solid curve and dots) derived from \mmpt\ data as in Sec.~\ref{tcmgeom} that does not reflect the consequences of parton time dilation. The trend may be compared  with those for \pp\ and \ppb\ collisions. Results in panels (a,b,c) reveal that the {\em effective} $\nu$ is \pt\ dependent, varying from values consistent with \mmpt\ data (and {\em integrated} hard components) at lower \pt\ to 1 at higher \pt. The dotted curve in (d) is the \ppb\ trend with constraint $\nu = 1$. The effective $\nu(p_t)$ trend for \pbpb\ at higher \pt\ is suggested by the solid triangles.

Note that \pp\ data hard components in ratio to TCM references (open circles) fall off at higher \pt. The same functional form (lowest solid curves) is exhibited for each hadron species. Given previous experience with TCM analysis of \pp\ spectra~\cite{alicetomspec,tomnewppspec} the falloff is likely an artifact. \pp\ data {\em in ratio} produce an {\em increasing} trend for $R_{AA}$ above 5 GeV/c as in Fig.~\ref{fig1} (b), whereas no corresponding trend is observed for $r_{Y}(y_t)$ in Fig.~\ref{fig6} (a,b) with  the same TCM reference.

As mentioned in Sec.~\ref{standard}, the spacing of \pbpb\ event classes on $\bar \rho_0$ is quite uneven as reflected by the results in Fig.~\ref{fig6}. The highest three event classes produce almost identical results in panels (a,b,c) and the lowest event class is well above the transition near $\bar \rho_0 = 15$,  as also indicated by the solid dots in panel (d). The transition is manifested in that panel by the point of separation among \pp, \ppb\ and \pbpb\ trends above which $N_{part}/2$ may deviate from 1.

%%%%%%%%%%%%%%%%%%
\section{Modified TCM -- p-Pb hard components} \label{ppb}

A major issue that has persisted since shortly after LHC startup is the possible formation of a QGP in smaller collision systems~\cite{alicenucmod} and even \pp\ collisions~\cite{cmsridge} based on apparent evidence emerging from some conventional analysis methods. One of the goals of the present and previous analyses has been resolution of that issue via study of 5 TeV \ppb\ spectra. In this section results from application of the modified TCM spectrum analysis to \ppb\ data is presented.

\begin{figure}[h]
	\includegraphics[width=1.46in,height=1.4in]{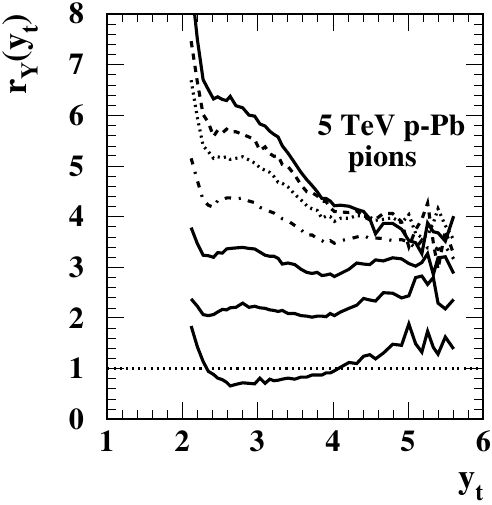}
	\includegraphics[width=1.46in,height=1.4in]{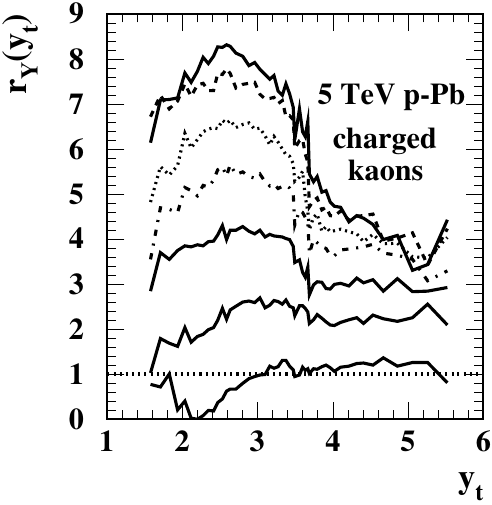}
	\includegraphics[width=2.92in,height=1.4in]{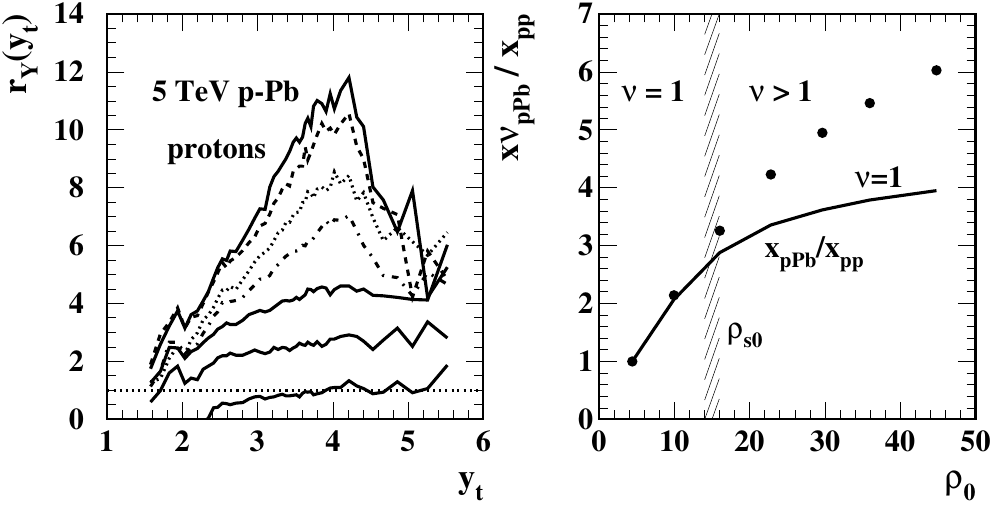}
	\caption{Revised spectrum ratio $r_Y(y_t)$ for \ppb\ collisions and for
		(a) pions,
		(b)  charged kaons and
		(c) protons;
		(d) ratio $x\nu_{pPb} / x_{pp}$ for \ppb\ collisions.
	}
	\label{fig7}     
\end{figure}

Figure~\ref{fig7} shows hard-component ratio $r_Y(y_t)$ for (a) pions, (b) charged kaons and (c) protons from seven event classes of 5 TeV \ppb\ collisions. These results appear to be substantially different from those in Fig.~\ref{fig6}. However, the apparent differences are mainly due to the very different spacing of event classes on event \nch. Referring to panel (d) the first three event classes are below or at the transition point $\bar \rho_0 = 15$ with $N_{part}/ 2 = 1$ to good approximation. The collision system is almost exclusively single \nn\ collisions and the hard-component shape on \pt\ remains approximately constant. For the last four event classes the increase at lower \pt\ follows the trend expected from \mmpt\ data (points in panel (d)), but stalls at and above \yt\ = 4, just as for \pbpb\ collisions in Fig.~\ref{fig6} (and as sketched by the curve for $\nu = 1$ in panel (d)).

%%%%%%%%%%
%%%%%%%%%%

This mismatch between \ppb\ and \pbpb\ event classes is curious. As shown in Fig.~\ref{fig2} (c) the \mmpt\ data for \ppb\ extend up to $\bar \rho_0 = 115$ whereas the density for the ``most central'' event class determined by Glauber Monte Carlo in Ref.~\cite{aliceglaub} is 45 (only a factor 3 above the transition). The \mmpt\ data for \pbpb\ finely covers $\bar \rho_0$ from 0 up to 175 yet the lowest two \pbpb\ event classes (solid squares) are at 55 (nearly factor 4 above the transition) and 205, neither overlapping the \ppb\ range. The result could be a frustrating lost opportunity to make a direct comparison between possible jet modification in small and large collision systems. However, because of the present highly-differential plot format and precise access to almost all jet fragments a useful comparison is possible. The same basic physics is manifested in both collision systems. Modest quantitative differences can be attributed to asymmetric vs symmetric collision geometries.

%%%%%%%%%%
\section{K/pi ratios and nuclear transparency} \label{kpi}

Figure~25 (upper panels) of Ref.~\cite{alicepbpbpidspec} shows K/$\pi$ spectrum ratios for the six centrality classes of \pbpb\ collisions considered in the present study. The paper observes that the spectrum ratios are approximately independent of \pt\ at ``high \pt'' (i.e.\ above 4 GeV/c) and approximately the same for \pbpb\ and \pp\ collisions. However, there is no explanation for that interesting result. The proton/pion ratio (Fig.\ 26) is more complex as should be expected due to substantial differences between meson and baryon jet fragment distributions~\cite{ppbpid}.
The \pt\ range emphasized in Fig.~25 corresponds to the interval $y_t > 4$ in Fig.~\ref{fig6} above. For pions and kaons, ratio $r_Y(y_t)$ is consistent with 1 for six centrality classes due to exclusivity and parton time dilation. The jet contribution to spectra does not change with \pbpb\ centrality for two meson species. Each of the six panels in Fig.~25 corresponds, within that \pt\ interval, in effect to single \nn\ collisions.

The TCM predicts a common ratio value in each panel. At higher \pt\ the hard component of Eq.~\ref{pidtcm} dominates. The ratio of two spectra for the relevant \yt\ interval is then approximately
\bea
\frac{\bar \rho_{0i}(y_t)}{\bar \rho_{0j}(y_t)} \rightarrow \frac{\tilde z_i}{\tilde z_j} \frac{\hat H_{0i}(y_t)}{\hat H_{0j}(y_t)}\frac{z_{0i}(1 + \tilde z_j x \nu)}{z_{0j}(1 + \tilde z_i x \nu)},
\eea
where $i$ is kaons and $j$ is pions. For the two meson species model functions $\hat H_0(y_t)$ cancel approximately. $\tilde z$ values for pions and kaons respectively are 0.6 and 2.6~\cite{pidpart1}. $z_0$ fractional abundances are 0.8 and 0.125 respectively, consistent with a statistical model~\cite{statmodel}. For  defined \pbpb\ event classes $x \approx 0.16$ and $\nu \approx 1$. That combination leads to constant $\bar \rho_{0i}(y_t)/\bar \rho_{0j}(y_t) \approx 0.5$ corresponding to Fig.~25 of Ref.~\cite{alicepbpbpidspec}. In effect, that figure confirms nuclear transparency for \pbpb\ collisions: for higher \pt\ each participant nucleon undergoes only a single \nn\ collision.

Below 4 GeV/c the spectrum ratios in Fig.~25 are quite deceptive. They can be contrasted with Fig.~\ref{fig6} (a,b) of the present study where the large jet contributions for pions and kaons are clearly evident within that \pt\ interval (i.e.\ $y_t < 4$). The spectrum ratios in Fig.~25 correspond to the ratio of panel (b) to panel (a) in Fig.~\ref{fig6}, where the similar lower-\pt\ jet contributions may cancel due to the similarity of two meson fragment distributions. Whereas an electrical engineer might welcome common-mode {\em noise} rejection, in this case one encounters common-mode {\em signal} rejection. In addition, ratios of {\em complete} spectra as in Fig.~25 overwhelm jet contributions at lower \pt\ with their included soft components. Thus, one observes only vestigial traces of jet contributions as small peaks near 2 GeV/c in the first few panels of Fig.~25.

%%%%%%%%%%
\section{Conclusions} \label{conclude}

Implications arising from these results are as follows: 
(a) Even for central \pbpb\ collisions the number of participant nucleons is about 1/3 that estimated by a classical Glauber Monte Carlo, in part because of exclusivity for \nn\ collisions as quantum transitions.
(b) Spectrum hard components do not factorize as is conventionally assumed (e.g.\ $R_\text{AA}$), in part because $N_{bin}$ or $\nu$ actually relates to individual projectile partons and depends then on time dilation corresponding to momentum fraction $x$. The {\em effective} value of $N_{bin}$ relating to produced hadrons is thus strongly \pt\ dependent.
(c) Hadron production corresponding to high-$x$ partons is restricted to single \nn\ collisions for corresponding participant nucleons, similar to {\em nuclear transparency} as observed for fixed-target \pa\ experiments in the seventies.
(d) The same analysis applied to \ppb\ spectra reveals  that a similar scenario applies, with small differences due to different collision geometries.
(e) A collision scenario involving formation of a high-density flowing QCD medium appears inconsistent with those findings.
It seems ironic that the apparently strongest diagnostic for jet suppression or quenching by a dense flowing QCD medium actually provides the strongest evidence for nuclear transparency (via exclusivity and time dilation).

%%%%%%%%%%%%%%%%%%%

%%%%%%%%%%%%%%%%%%%%%

%%%%%%%%%%%%%%%%%%%

\end{document}